\documentclass[prl,aps,
twocolumn,
nofootinbib,
floatfix,
preprintnumbers,
superscriptaddress]{revtex4}

\usepackage{hyperref}
\usepackage[dvipsnames]{xcolor}
\usepackage{amsmath,amssymb,mathrsfs,amsthm,tikz,shuffle,paralist,accents}

\usetikzlibrary{shapes.misc}
\usetikzlibrary{cd}
\usetikzlibrary{snakes}
\usetikzlibrary{calc}
\usetikzlibrary{fadings}

\usepackage[all]{xy}

\tikzset{
    ncbar angle/.initial=90,
    ncbar/.style={
        to path=(\tikztostart)
        -- ($(\tikztostart)!#1!\pgfkeysvalueof{/tikz/ncbar angle}:(\tikztotarget)$)
        -- ($(\tikztotarget)!($(\tikztostart)!#1!\pgfkeysvalueof{/tikz/ncbar angle}:(\tikztotarget)$)!\pgfkeysvalueof{/tikz/ncbar angle}:(\tikztostart)$)
        -- (\tikztotarget)
    },
    ncbar/.default=0.5cm,
}
\tikzset{round left paren/.style={ncbar=0.5cm,out=115,in=-115}}
\tikzset{round right paren/.style={ncbar=0.5cm,out=60,in=-60}}

\begin{document}

\preprint{CERN-TH-2023-104}

\title{Traintracks All the Way Down}

\author{Andrew~J.~McLeod}
\affiliation{CERN, Theoretical Physics Department, 1211 Geneva 23, Switzerland}
\affiliation{Mani L. Bhaumik Institute for Theoretical Physics, Department of Physics and Astronomy,
UCLA, Los Angeles, CA 90095, USA}
\author{Matt~von~Hippel}
\affiliation{Niels Bohr International Academy and Discovery Center, Niels Bohr Institute\\
University of Copenhagen, Blegdamsvej 17, DK-2100, Copenhagen \O, Denmark}

\begin{abstract}
We study the class of planar Feynman integrals that can be constructed by sequentially intersecting traintrack diagrams without forming a closed traintrack loop. After describing how to derive a $2L$-fold integral representation of any $L$-loop diagram in this class, we provide evidence that their leading singularities always give rise to integrals over $(L{-}1)$-dimensional varieties for generic external momenta, which for certain graphs we can identify as Calabi-Yau $(L{-}1)$-folds. We then show that these diagrams possess an interesting nested structure, due to the large number of second-order differential operators that map them to (products of) lower-loop integrals of the same type. 
\end{abstract}

\ \vspace{.1cm}
\maketitle

\noindent {\bf Introduction}
\vspace{.1cm}

A great deal can be learned about quantum field theory by identifying ways in which it exhibits recursive structure. In fact, some of the most striking advances in our ability to compute scattering amplitudes have come from harnessing such structure, as done for instance by the Berends-Giele and BCFW recursions~\cite{Berends:1987me,Britto:2004ap,Britto:2005fq}, the MHV vertex expansion~\cite{Cachazo:2004kj}, and the recursive construction of all-loop amplitude integrands in the planar limit of maximally supersymmetric gauge theory~\cite{Arkani-Hamed:2010zjl}. By providing constructive solutions to problems that were previously computationally intractable, these kinds of recursions allow us to ask new types of questions that probe ever more deeply into the structure of quantum field theory. 

In like manner, the recent identification of several classes of Feynman diagrams that give rise to integrals over varieties whose dimension grows with the loop order~\cite{Bloch:2014qca,Bloch:2016izu,Bourjaily:2018ycu,Bourjaily:2018yfy,Duhr:2022pch,Duhr:2022dxb} has opened up new and interesting avenues of study. Some of the chief questions that have been raised include whether or not these Feynman integral varieties exhibit any universal geometric properties, whether these geometries can be seen to encode any basic physical principles, and how we can go about gaining greater analytic and numerical control over these types of integrals.

While no definitive answers to these questions have yet been forthcoming, important progress has been made on a number of fronts~\cite{Bourjaily:2022bwx}. Much of this progress has come from the ongoing study of the massive sunrise diagrams (see~\cite{Klemm:2019dbm,Bonisch:2020qmm,Bonisch:2021yfw,Pogel:2022yat,Duhr:2022dxb,Forum:2022lpz,Pogel:2022ken,Pogel:2022vat} for some recent developments), which are known to give rise to integrals over Calabi-Yau $(L{-}1)$-folds at $L$ loops~\cite{Bloch:2014qca,Bloch:2016izu}. The varieties that appear in massless diagrams such as the traintrack and tardigrade diagrams~\cite{Bourjaily:2018ycu,Bourjaily:2018yfy} have also been characterized with increasing rigour~\cite{Bourjaily:2019hmc,Lairez:2022zkj,Duhr:2022pch,McLeod:2023qdf,Cao:2023tpx,Doran:2023yzu}. However, even with these advances, we remain in a situation where the bulk of what we know about the special functions that can appear in perturbative quantum field theory comes from the study of a relatively restricted set of diagrams.

\begin{figure}[b]
\centering
\begin{tikzpicture}[scale=0.66]
\node (d11) at (-4,.5) {};
\node (d12) at (-3,.5) {};
\node (d13) at (-2,.5) {};
\node (d14) at (-1,.5) {};
\node (d15) at (0,.5) {};
\node (d16) at (0,-.5) {};
\node (d17) at (-1,-.5) {};
\node (d18)at (-2,-.5) {};
\node (d19) at (-3,-.5) {};
\node (d110) at (-4,-.5) {};
\node (d111) at (-2,-1.5) {};
\node (d112) at (-3,-1.5) {};
\node (d113) at (-1,1.5) {};
\node (d114) at (-1,2.5) {};
\node (d115) at (-1,3.5) {};
\node (d116) at (-2,3.5) {};
\node (d117) at (-2,2.5) {};
\node (d118) at (-2,1.5) {};
\node (d119) at (-3,2.5) {};
\node (d120) at (-3,1.5) {};
\node (d121) at (0,3.5) {};
\node (d122) at (1,3.5) {};
\node (d123) at (1,2.5) {};
\node (d124) at (0,2.5) {};
\draw[very thick]  (d11.center) edge (d12.center);
\draw[very thick]  (d12.center) edge (d13.center);
\draw[very thick]  (d13.center) edge (d14.center);
\draw[very thick]  (d14.center) edge (d15.center);
\draw[very thick]  (d15.center) edge (d16.center);
\draw[very thick]  (d16.center) edge (d17.center);
\draw[very thick]  (d17.center) edge (d18.center);
\draw[very thick]  (d18.center) edge (d19.center);
\draw[very thick]  (d19.center) edge (d110.center);`
\draw[very thick]  (d110.center) edge (d11.center);
\draw[very thick]  (d18.center) edge (d111.center);
\draw[very thick]  (d111.center) edge (d112.center);
\draw[very thick]  (d112.center) edge (d19.center);
\draw[very thick]  (d14.center) edge (d113.center);
\draw[very thick]  (d113.center) edge (d114.center);
\draw[very thick]  (d114.center) edge (d115.center);
\draw[very thick]  (d115.center) edge (d116.center);
\draw[very thick]  (d116.center) edge (d117.center);
\draw[very thick]  (d117.center) edge (d118.center);
\draw[very thick]  (d118.center) edge (d13.center);
\draw[very thick]  (d117.center) edge (d119.center);
\draw[very thick]  (d119.center) edge (d120.center);
\draw[very thick]  (d120.center) edge (d118.center);
\draw[very thick]  (d115.center) edge (d121.center);
\draw[very thick]  (d121.center) edge (d122.center);
\draw[very thick]  (d122.center) edge (d123.center);
\draw[very thick]  (d123.center) edge (d124.center);
\draw[very thick]  (d124.center) edge (d114.center);
\draw[very thick]  (d12.center) edge (d19.center);
\draw[very thick]  (d13.center) edge (d18.center);
\draw[very thick]  (d14.center) edge (d17.center);
\draw[very thick]  (d113.center) edge (d118.center);
\draw[very thick]  (d114.center) edge (d117.center);
\draw[very thick]  (d119.center) edge (d120.center);
\draw[very thick]  (d121.center) edge (d124.center);
\draw[very thick]  (d11.center) edge +(0,.3);
\draw[very thick]  (d11.center) edge +(-.3,0);
\draw[very thick]  (d15.center) edge +(0,.3);
\draw[very thick]  (d15.center) edge +(.3,0);
\draw[very thick]  (d16.center) edge +(0,-.3);
\draw[very thick]  (d16.center) edge +(.3,0);
\draw[very thick]  (d110.center) edge +(0,-.3);
\draw[very thick]  (d110.center) edge +(-.3,0);
\draw[very thick]  (d111.center) edge +(0,-.3);
\draw[very thick]  (d111.center) edge +(.3,0);
\draw[very thick]  (d112.center) edge +(0,-.3);
\draw[very thick]  (d112.center) edge +(-.3,0);
\draw[very thick]  (d116.center) edge +(0,.3);
\draw[very thick]  (d116.center) edge +(-.3,0);
\draw[very thick]  (d119.center) edge +(0,.3);
\draw[very thick]  (d119.center) edge +(-.3,0);
\draw[very thick]  (d120.center) edge +(0,-.3);
\draw[very thick]  (d120.center) edge +(-.3,0);
\draw[very thick]  (d122.center) edge +(0,.3);
\draw[very thick]  (d122.center) edge +(.3,0);
\draw[very thick]  (d123.center) edge +(0,-.3);
\draw[very thick]  (d123.center) edge +(.3,0);
\draw[very thick]  (d115.center) edge +(0,.3);
\draw[very thick]  (d121.center) edge +(.1,.3);
\draw[very thick]  (d121.center) edge +(-.1,.3);
\draw[very thick]  (d113.center) edge +(.3,0);
\draw[very thick]  (d114.center) edge +(.2,-.3);
\draw[very thick]  (d114.center) edge +(.3,-.2);
\node (d21) at (1.5,-1.25) {};
\node (d22) at (1.5,-.25) {};
\node (d23) at (1.5,.75) {};
\node (d24) at (2.5,.75) {};
\node (d25) at (2.5,-.25) {};
\node (d26) at (2.5,-1.25) {};
\node (d27) at (2.5,-2.25) {};
\node (d28) at (1.5,-2.25) {};
\node (d29) at (0.5,-2.25) {};
\node (d210) at (-.5,-2.25) {};
\node (d211) at (-.5,-1.25) {};
\node (d212) at (0.5,-1.25) {};
\node (d213) at (3.5,-1.25) {};
\node (d214) at (3.5,-2.25) {};
\node (d215) at (3.5,-.25) {};
\node (d216) at (3.5,.75) {};
\node (d217) at (3.5,1.75) {};
\node (d218) at (3.5,2.75) {};
\node (d219) at (2.5,2.75) {};
\node (d220) at (2.5,1.75) {};
\draw[very thick]  (d21.center) edge (d22.center);
\draw[very thick]  (d22.center) edge (d23.center);
\draw[very thick]  (d23.center) edge (d24.center);
\draw[very thick]  (d24.center) edge (d25.center);
\draw[very thick]  (d25.center) edge (d26.center);
\draw[very thick]  (d26.center) edge (d27.center);
\draw[very thick]  (d27.center) edge (d28.center);
\draw[very thick]  (d28.center) edge (d29.center);
\draw[very thick]  (d29.center) edge (d210.center);
\draw[very thick]  (d210.center) edge (d211.center);
\draw[very thick]  (d211.center) edge (d212.center);
\draw[very thick]  (d212.center) edge (d21.center);
\draw[very thick]  (d26.center) edge (d213.center);
\draw[very thick]  (d213.center) edge (d214.center);
\draw[very thick]  (d214.center) edge (d27.center);
\draw[very thick]  (d25.center) edge (d215.center);
\draw[very thick]  (d215.center) edge (d216.center);
\draw[very thick]  (d216.center) edge (d217.center);
\draw[very thick]  (d217.center) edge (d218.center);
\draw[very thick]  (d218.center) edge (d219.center);
\draw[very thick]  (d219.center) edge (d220.center);
\draw[very thick]  (d220.center) edge (d24.center);
\draw[very thick]  (d21.center) edge (d26.center);
\draw[very thick]  (d22.center) edge (d25.center);
\draw[very thick]  (d21.center) edge (d28.center);
\draw[very thick]  (d212.center) edge (d29.center);
\draw[very thick]  (d220.center) edge (d217.center);
\draw[very thick]  (d24.center) edge (d216.center);
\draw[very thick]  (d211.center) edge +(0,.3);
\draw[very thick]  (d211.center) edge +(-.3,0);
\draw[very thick]  (d210.center) edge +(0,-.3);
\draw[very thick]  (d210.center) edge +(-.3,0);
\draw[very thick]  (d213.center) edge +(0,.3);
\draw[very thick]  (d213.center) edge +(.3,0);
\draw[very thick]  (d214.center) edge +(0,-.3);
\draw[very thick]  (d214.center) edge +(.3,0);
\draw[very thick]  (d23.center) edge +(0,.3);
\draw[very thick]  (d23.center) edge +(-.3,0);
\draw[very thick]  (d218.center) edge +(0,.3);
\draw[very thick]  (d218.center) edge +(.3,0);
\draw[very thick]  (d219.center) edge +(0,.3);
\draw[very thick]  (d219.center) edge +(-.3,0);
\draw[very thick]  (d215.center) edge +(0,-.3);
\draw[very thick]  (d215.center) edge +(.3,0);
\draw[very thick]  (d220.center) edge +(-.3,.1);
\draw[very thick]  (d220.center) edge +(-.3,-.1);
\draw[very thick]  (d217.center) edge +(.3,.1);
\draw[very thick]  (d217.center) edge +(.3,-.1);
\draw[very thick]  (d27.center) edge +(-.1,-.3);
\draw[very thick]  (d27.center) edge +(.1,-.3);
\draw[very thick]  (d28.center) edge +(0,-.3);
\draw[very thick]  (d29.center) edge +(-.1,-.3);
\draw[very thick]  (d29.center) edge +(.1,-.3);
\node (d31) at (5.75,-2) {};
\node (d32) at (5.75,-1) {};
\node (d33) at (5.75,0) {};
\node (d34) at (5.75,1) {};
\node (d35) at (5.75,2) {};
\node (d36) at (5.75,3) {};
\node (d37) at (5.75,4) {};
\node (d38) at (6.75,4) {};
\node (d39) at (6.75,3) {};
\node (d310) at (6.75,2) {};
\node (d311) at (6.75,1) {};
\node (d312) at (6.75,0) {};
\node (d313) at (6.75,-1) {};
\node (d314) at (6.75,-2) {};
\node (d315) at (4.75,-1) {};
\node (d316) at (4.75,0) {};
\node (d317) at (4.75,2) {};
\node (d318) at (4.75,3) {};
\node (d319) at (4.75,3) {};
\node (d320) at (4.75,4) {};
\node (d321) at (7.75,3) {};
\node (d322) at (7.75,2) {};
\node (d323) at (7.75,1) {};
\node (d324) at (7.75,0) {};
\draw[very thick]  (d31.center) edge (d32.center);
\draw[very thick]  (d32.center) edge (d33.center);
\draw[very thick]  (d33.center) edge (d34.center);
\draw[very thick]  (d34.center) edge (d35.center);
\draw[very thick]  (d35.center) edge (d36.center);
\draw[very thick]  (d36.center) edge (d37.center);
\draw[very thick]  (d37.center) edge (d38.center);
\draw[very thick]  (d38.center) edge (d39.center);
\draw[very thick]  (d39.center) edge (d310.center);
\draw[very thick]  (d310.center) edge (d311.center);
\draw[very thick]  (d311.center) edge (d312.center);
\draw[very thick]  (d312.center) edge (d313.center);
\draw[very thick]  (d313.center) edge (d314.center);
\draw[very thick]  (d314.center) edge (d31.center);
\draw[very thick]  (d32.center) edge (d315.center);
\draw[very thick]  (d315.center) edge (d316.center);
\draw[very thick]  (d316.center) edge (d33.center);
\draw[very thick]  (d35.center) edge (d317.center);
\draw[very thick]  (d317.center) edge (d318.center);
\draw[very thick]  (d318.center) edge (d36.center);
\draw[very thick]  (d39.center) edge (d321.center);
\draw[very thick]  (d321.center) edge (d322.center);
\draw[very thick]  (d322.center) edge (d310.center);
\draw[very thick]  (d311.center) edge (d323.center);
\draw[very thick]  (d323.center) edge (d324.center);
\draw[very thick]  (d324.center) edge (d312.center);
\draw[very thick]  (d32.center) edge (d313.center);
\draw[very thick]  (d33.center) edge (d312.center);
\draw[very thick]  (d34.center) edge (d311.center);
\draw[very thick]  (d35.center) edge (d310.center);
\draw[very thick]  (d36.center) edge (d39.center);
\draw[very thick]  (d31.center) edge +(0,-.3);
\draw[very thick]  (d31.center) edge +(-.3,0);
\draw[very thick]  (d314.center) edge +(0,-.3);
\draw[very thick]  (d314.center) edge +(.3,0);
\draw[very thick]  (d315.center) edge +(0,-.3);
\draw[very thick]  (d315.center) edge +(-.3,0);
\draw[very thick]  (d316.center) edge +(0,.3);
\draw[very thick]  (d316.center) edge +(-.3,0);
\draw[very thick]  (d317.center) edge +(0,-.3);
\draw[very thick]  (d317.center) edge +(-.3,0);
\draw[very thick]  (d318.center) edge +(0,.3);
\draw[very thick]  (d318.center) edge +(-.3,0);
\draw[very thick]  (d321.center) edge +(0,.3);
\draw[very thick]  (d321.center) edge +(.3,0);
\draw[very thick]  (d322.center) edge +(0,-.3);
\draw[very thick]  (d322.center) edge +(.3,0);
\draw[very thick]  (d323.center) edge +(0,.3);
\draw[very thick]  (d323.center) edge +(.3,0);
\draw[very thick]  (d324.center) edge +(0,-.3);
\draw[very thick]  (d324.center) edge +(.3,0);
\draw[very thick]  (d32.center) edge +(-.25,-.25);
\draw[very thick]  (d34.center) edge +(-.3,0);
\draw[very thick]  (d35.center) edge +(-.2,-.3);
\draw[very thick]  (d35.center) edge +(-.3,-.2);
\draw[very thick]  (d36.center) edge +(-.2,.3);
\draw[very thick]  (d36.center) edge +(-.3,.2);
\draw[very thick]  (d37.center) edge +(0,.3);
\draw[very thick]  (d37.center) edge +(-.3,0);
\draw[very thick]  (d38.center) edge +(0,.3);
\draw[very thick]  (d38.center) edge +(.3,0);
\draw[very thick]  (d39.center) edge +(.2,.3);
\draw[very thick]  (d39.center) edge +(.3,.2);
\draw[very thick]  (d310.center) edge +(.2,-.3);
\draw[very thick]  (d310.center) edge +(.3,-.2);
\draw[very thick]  (d312.center) edge +(.25,-.25);
\draw[very thick]  (d313.center) edge +(.3,0);
\end{tikzpicture} 
\caption{Three examples of traintrack network diagrams. By attaching a pair of massless external lines to each vertex with just two internal lines, these diagrams are rendered infrared-finite. One can additionally assign one or more external momenta to each other vertex.}
\label{fig:example_traintrack_networks}
\end{figure}
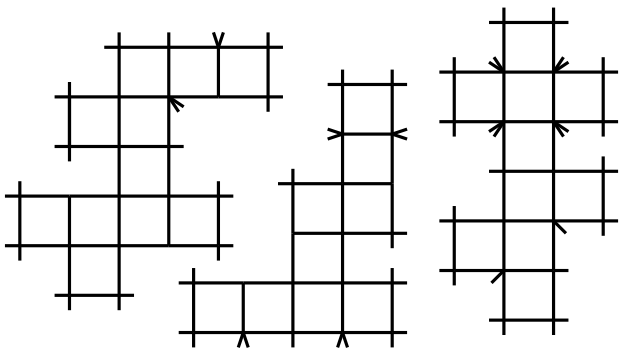

In this letter, we enlarge the scope of these investigations into the varieties that appear in Feynman integrals to cover a new class of diagrams that generalize the traintrack diagrams. In particular, the diagrams we consider are those that can be constructed by iteratively intersecting traintrack diagrams, where we exclude only diagrams that involve closed traintrack loops (or, whose dual diagrams are not trees). We call these diagrams \emph{traintrack network diagrams} (or simply traintrack networks), while for clarity we refer to the original traintrack diagrams as \emph{linear traintrack diagrams}.  Three examples of traintrack networks are shown in Figure~\ref{fig:example_traintrack_networks}.

To study the types of varieties that appear in traintrack networks, we first describe a recursive algorithm for deriving a $2L$-fold Feynman-parameter integral representation of any diagram in this class. We then show that one can always compute $L{+}1$ residues of the integrand, and that---for generic kinematics---this results in an integral over an $(L{-}1)$-dimensional variety. This matches the dimension found for the linear traintrack diagrams~\cite{Bourjaily:2018ycu,Cao:2023tpx}; however, beyond a few special cases, it is not immediately clear whether the varieties that appear in traintrack networks are always Calabi-Yau, due to the appearance of nontrivial numerator factors. 

We also study aspects of the analytic structure of traintrack networks that go beyond their leading singularities, by showing that these diagrams satisfy a large web of second-order differential equations that relate integrals---or products of integrals---at adjacent loop orders. These differential equations allow us to concretely identify ways in which the integral varieties associated with subtopologies enter higher-loop traintrack networks, and provide us with constraints that we expect will prove useful for bootstrapping these integrals in the future. More generally, the nested structure that these differential equations expose hints at the existence of a different formulation of traintrack networks that would make clear the types of special functions they evaluate to at high loop orders.

\vspace{.2cm}
\noindent {\bf Traintrack Networks}
\vspace{.1cm}

To explore the properties of traintrack diagrams, we begin by defining our notation. Similar to linear traintrack diagrams, traintrack networks can be made dual conformal invariant through the inclusion of an appropriate numerator. To make this symmetry manifest, we express the momentum flowing through each edge as the difference between a pair of dual points associated with the faces that border it in the graph. For instance, the momentum $p$ flowing through an edge that borders faces $a$ and $b$ of the Feynman diagram becomes the difference $p = x_b - x_a$. In terms of these variables, momentum space propagators become (inverse) squared differences of dual points, which we denote by $(a,b) = (x_b - x_a)^2$, while integrals over loop momentum become integrals over the position of internal dual points.\footnote{This notation is intended to emphasize that $(a,b)$ is a bilinear inner product in embedding space (see for example~\cite{Hodges:2010kq,Simmons-Duffin:2012juh,Bartels:2014mka,Bourjaily:2019exo}).}  

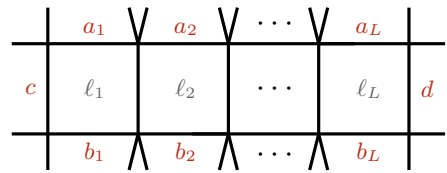
\begin{figure}[t]
\centering
\begin{tikzpicture}[scale=1.2]
\node (d11) at (-4,.5) {};
\node (d12) at (-3,.5) {};
\node (d13) at (-2,.5) {};
\node (d14) at (-1,.5) {};
\node (d15) at (0,.5) {};
\node (d16) at (0,-.5) {};
\node (d17) at (-1,-.5) {};
\node (d18)at (-2,-.5) {};
\node (d19) at (-3,-.5) {};
\node (d110) at (-4,-.5) {};
\draw[very thick]  (d11.center) edge (d12.center);
\draw[very thick]  (d12.center) edge (d13.center);
\draw[very thick]  (d13.center) edge (d14.center);
\draw[very thick]  (d14.center) edge (d15.center);
\draw[very thick]  (d15.center) edge (d16.center);
\draw[very thick]  (d16.center) edge (d17.center);
\draw[very thick]  (d17.center) edge (d18.center);
\draw[very thick]  (d18.center) edge (d19.center);
\draw[very thick]  (d19.center) edge (d110.center);
\draw[very thick]  (d110.center) edge (d11.center);
\draw[very thick]  (d12.center) edge (d19.center);
\draw[very thick]  (d13.center) edge (d18.center);
\draw[very thick]  (d14.center) edge (d17.center);
\draw[very thick]  (d11.center) edge +(-.4,0);
\draw[very thick]  (d11.center) edge +(0,.4);
\draw[very thick]  (d14.center) edge +(.4,0);
\draw[very thick]  (d15.center) edge +(.4,0);
\draw[very thick]  (d15.center) edge +(0,.4);
\draw[very thick]  (d18.center) edge +(-.4,0);
\draw[very thick]  (d12.center) edge +(-.1,.4);
\draw[very thick]  (d12.center) edge +(.1,.4);
\draw[very thick]  (d13.center) edge +(-.1,.4);
\draw[very thick]  (d13.center) edge +(.1,.4);
\draw[very thick]  (d16.center) edge +(0,-.4);
\draw[very thick]  (d16.center) edge +(.4,0);
\draw[very thick]  (d17.center) edge +(-.1,-.4);
\draw[very thick]  (d17.center) edge +(.1,-.4);
\draw[very thick]  (d14.center) edge +(-.1,.4);
\draw[very thick]  (d14.center) edge +(.1,.4);
\draw[very thick]  (d18.center) edge +(0.1,-.4);
\draw[very thick]  (d18.center) edge +(-.1,-.4);
\draw[very thick]  (d19.center) edge +(0.1,-.4);
\draw[very thick]  (d19.center) edge +(-.1,-.4);
\draw[very thick]  (d110.center) edge +(0,-.4);
\draw[very thick]  (d110.center) edge +(-.4,0);
\node[label=left:{${\color{BrickRed} c}$}] at (-3.94,0) {};
\node[label=right:{${\color{BrickRed} d}$}] at (-.06,0) {};
\node[label=above:{${\color{BrickRed} a_1}$}] at (-3.48,.4) {};
\node[label=above:{${\color{BrickRed} a_2}$}] at (-2.47,.4) {};
\node[label=above:{${\color{BrickRed} a_L}$}] at (-.44,.4) {};
\node[label=above:{\large $\cdots$}] at (-1.46,.45) {};
\node[label=center:{\large $\cdots$}] at (-1.46,0) {};
\node[label=below:{${\color{BrickRed} b_1}$}] at (-3.48,-.4) {};
\node[label=below:{${\color{BrickRed} b_2}$}] at (-2.47,-.4) {};
\node[label=below:{${\color{BrickRed} b_L}$}] at (-.44,-.4) {};
\node[label=below:{\large $\cdots$}] at (-1.46,-.46) {};
\node[label=center:{$\color{black!60} \ell_{1}$}] at (-3.48,0) {};
\node[label=center:{$\color{black!60} \ell_{2}$}] at (-2.47,0) {};
\node[label=center:{$\color{black!60} \ell_{L}$}] at (-.44,0) {};
\end{tikzpicture} 
\caption{The linear traintrack diagram, which depends on the external dual points $({\color{BrickRed} a_i})$, $({\color{BrickRed} b_i})$, $({\color{BrickRed} c})$ and $({\color{BrickRed} d})$.}
\label{fig:linear_traintrack}
\end{figure}

Traintrack networks that involve intersecting a large number of linear traintrack diagrams will in general involve intricate patterns of propagators. However, as we now show, a $2L$-fold integral representation can always be derived for traintrack networks by iteratively applying the loop-by-loop Feynman parametrization described in~\cite{Bourjaily:2018ycu} (see also~\cite{Bourjaily:2019jrk,Bourjaily:2018aeq,Cao:2023tpx}). Briefly, we recall that this method allows one to express the linear $L$-loop traintrack diagram depicted in Figure~\ref{fig:linear_traintrack} as the integral
\begin{equation}
\int_0^\infty \frac{d^L {\color{ForestGreen} \alpha} \, d^L {\color{ForestGreen} \beta} \, d^L {\color{ForestGreen} \gamma}}{\text{GL}(1)^L}  \frac{\mathcal{N}}{[({\color{Cerulean} R_1},{\color{Cerulean} R_1}) \cdots ({\color{Cerulean} R_L}, {\color{Cerulean} R_L})] ({\color{Cerulean} R_L}, {\color{BrickRed} d})} \, , \label{eq:traintrack_parametric_form}
\end{equation}
where $\mathcal{N}$ is the numerator of the original Feynman diagram, and the $\text{GL}(1)$ redundancies represent the fact that this integral is projective in every triplet of integration variables $\{{\color{ForestGreen} \alpha_i}, {\color{ForestGreen} \beta_i}, {\color{ForestGreen} \gamma_i}\}$. The composite dual points that appear in these integrals correspond to the linear combinations
\begin{equation}
\qquad ({\color{Cerulean} R_i}) = {\color{ForestGreen} \gamma_i} ({\color{Cerulean} R_{i-1}}) + {\color{ForestGreen} \alpha_i} ({\color{BrickRed} a_i}) + {\color{ForestGreen} \beta_i} ({\color{BrickRed} b_i}) \, ,
\end{equation}
where $({\color{Cerulean} R_0}) = ({\color{BrickRed} c})$. After fixing the projective redundancy of each triplet $\{{\color{ForestGreen} \alpha_i}, {\color{ForestGreen} \beta_i}, {\color{ForestGreen} \gamma_i}\}$ (for instance by setting ${\color{ForestGreen} \beta_i}=1$), this provides us with a manifestly $2L$-fold integral representation of the linear traintrack diagrams.

\begin{figure}
\centering
\begin{tikzpicture}[scale=0.9]
\node (d11) at (-4,.5) {};
\node (d12) at (-3,.5) {};
\node (d13) at (-2,.5) {};
\node (d14) at (-1,.5) {};
\node (d15) at (-1,-.5) {};
\node (d16)at (-2,-.5) {};
\node (d17) at (-3,-.5) {};
\node (d18) at (-4,-.5) {};
\draw[very thick]  (d11.center) edge (d12.center);
\draw[very thick]  (d12.center) edge (d13.center);
\draw[very thick]  (d13.center) edge (d14.center);
\draw[very thick]  (d14.center) edge (d15.center);
\draw[very thick]  (d15.center) edge (d16.center);
\draw[very thick]  (d16.center) edge (d17.center);
\draw[very thick]  (d17.center) edge (d18.center);
\draw[very thick]  (d18.center) edge (d11.center);
\draw[very thick]  (d12.center) edge (d17.center);
\draw[very thick]  (d13.center) edge (d16.center);
\draw[very thick]  (d11.center) edge +(-.4,0);
\draw[very thick]  (d14.center) edge +(.4,0);
\draw[very thick]  (d15.center) edge +(.4,0);
\draw[very thick]  (d18.center) edge +(-.4,0);
\draw[very thick]  (d12.center) edge +(-.1,.4);
\draw[very thick]  (d12.center) edge +(.1,.4);
\draw[very thick]  (d13.center) edge +(-.1,.4);
\draw[very thick]  (d13.center) edge +(.1,.4);
\draw[very thick]  (d16.center) edge +(-.1,-.4);
\draw[very thick]  (d16.center) edge +(.1,-.4);
\draw[very thick]  (d17.center) edge +(-.1,-.4);
\draw[very thick]  (d17.center) edge +(.1,-.4);
\draw[very thick]  (d11.center) edge +(-.1,.4);
\draw[very thick]  (d11.center) edge +(.1,.4);
\draw[very thick]  (d14.center) edge +(-.1,.4);
\draw[very thick]  (d14.center) edge +(.1,.4);
\draw[very thick]  (d15.center) edge +(-.1,-.4);
\draw[very thick]  (d15.center) edge +(.1,-.4);
\draw[very thick]  (d18.center) edge +(-.1,-.4);
\draw[very thick]  (d18.center) edge +(.1,-.4);
\node[label=left:{$\color{BrickRed} \cdots$}] at (-3.8,0) {};
\node[label=left:{$\color{BrickRed} \cdots$}] at (-.12,0) {};
\node[label=above:{$\color{BrickRed} a_{i-1}$}] at (-3.48,.4) {};
\node[label=above:{$\color{Cerulean} R_L^\prime$}] at (-2.47,.4) {};
\node[label=above:{$\color{BrickRed} a_{i+1}$}] at (-1.46,.4) {};
\node[label=below:{$\color{BrickRed} b_{i{-}1}$}] at (-3.48,-.4) {};
\node[label=below:{$\color{Cerulean} R_L^{\prime \prime}$}] at (-2.47,-.4) {};
\node[label=below:{$\color{BrickRed} b_{i+1}$}] at (-1.46,-.4) {};
\node[label=left:{$\color{black!60} \ell_{i-1}$}] at (-2.9,0) {};
\node[label=left:{$\color{black!60} \ell_i$}] at (-2.03,0) {};
\node[label=left:{$\color{black!60} \ell_{i+1}$}] at (-.92,0) {};
\node (d21) at (-3,3.5) {};
\node (d22) at (-2,3.5) {};
\node (d23) at (-2,2.5) {};
\node (d24) at (-3,2.5) {};
\draw[very thick]  (d21.center) edge (d22.center);
\draw[very thick]  (d22.center) edge (d23.center);
\draw[very thick]  (d23.center) edge (d24.center);
\draw[very thick]  (d24.center) edge (d21.center);
\draw[very thick]  (d21.center) edge +(0,.4);
\draw[very thick]  (d22.center) edge +(0,.4);
\draw[very thick]  (d23.center) edge +(0,-.4);
\draw[very thick]  (d23.center) edge +(.4,0);
\draw[very thick]  (d24.center) edge +(0,-.4);
\draw[very thick]  (d24.center) edge +(-.4,0);
\draw[very thick]  (d22.center) edge +(.4,.1);
\draw[very thick]  (d22.center) edge +(.4,-.1);
\draw[very thick]  (d21.center) edge +(-.4,.1);
\draw[very thick]  (d21.center) edge +(-.4,-.1);
\node[label=below:{$\color{BrickRed} \vdots$}] at (-2.5,4.5) {};
\node[label=below:{$\color{black!60} \ell_i$}] at (-2.47,2.6) {};
\node[label=left:{$\color{BrickRed} b_L^\prime$}] at (-2.9,2.95) {};
\node[label=right:{$\color{BrickRed} a_L^\prime$}] at (-2.1,2.95) {};
\node[label=left:{$\color{black!60} \ell_L^\prime$}] at (-2.03,3) {};
\node (d31) at (-3,-2.5) {};
\node (d32) at (-2,-2.5) {};
\node (d33) at (-2,-3.5) {};
\node (d34) at (-3,-3.5) {};
\draw[very thick]  (d31.center) edge (d32.center);
\draw[very thick]  (d32.center) edge (d33.center);
\draw[very thick]  (d33.center) edge (d34.center);
\draw[very thick]  (d34.center) edge (d31.center);
\draw[very thick]  (d31.center) edge +(0,.4);
\draw[very thick]  (d31.center) edge +(-.4,0);
\draw[very thick]  (d32.center) edge +(0,.4);
\draw[very thick]  (d32.center) edge +(.4,0);
\draw[very thick]  (d33.center) edge +(0,-.4);
\draw[very thick]  (d34.center) edge +(0,-.4);
\draw[very thick]  (d33.center) edge +(.4,.1);
\draw[very thick]  (d33.center) edge +(.4,-.1);
\draw[very thick]  (d34.center) edge +(-.4,.1);
\draw[very thick]  (d34.center) edge +(-.4,-.1);
\node[label=below:{$\color{BrickRed} \vdots$}] at (-2.5,-3.2) {};
\node[label=above:{$\color{black!60} \ell_i$}] at (-2.47,-2.6) {};
\node[label=left:{$\color{BrickRed} a_L^{\prime \prime}$}] at (-2.9,-2.95) {};
\node[label=right:{$\color{BrickRed} b_L^{\prime \prime}$}] at (-2.1,-2.95) {};
\node[label=left:{$\color{black!60} \ell_L^{\prime \prime}$}] at (-2.03,-3) {};
\draw [very thick, ->] (-2.5,1.8) -- (-2.5,1.25);
\draw [very thick, ->] (-2.5,-1.8) -- (-2.5,-1.25);
\draw [very thick,{Implies}-{Implies},double] (-.2,0) -- (.4,0);
\node (d41) at (1.2,.5) {};
\node (d42) at (2.3,.5) {};
\node (d43) at (3.3,.5) {};
\node (d44) at (4.4,.5) {};
\node (d45) at (4.4,-.5) {};
\node (d46)at (3.3,-.5) {};
\node (d47) at (2.3,-.5) {};
\node (d48) at (1.2,-.5) {};
\node (d49) at (2.3,1.6) {};
\node (d410) at (3.3,1.6) {};
\node (d411) at (3.3,-1.6) {};
\node (d412) at (2.3,-1.6) {};
\draw[very thick]  (d41.center) edge (d42.center);
\draw[very thick]  (d42.center) edge (d43.center);
\draw[very thick]  (d43.center) edge (d44.center);
\draw[very thick]  (d44.center) edge (d45.center);
\draw[very thick]  (d45.center) edge (d46.center);
\draw[very thick]  (d46.center) edge (d47.center);
\draw[very thick]  (d47.center) edge (d48.center);
\draw[very thick]  (d48.center) edge (d41.center);
\draw[very thick]  (d42.center) edge (d49.center);
\draw[very thick]  (d49.center) edge (d410.center);
\draw[very thick]  (d410.center) edge (d43.center);
\draw[very thick]  (d46.center) edge (d411.center);
\draw[very thick]  (d411.center) edge (d412.center);
\draw[very thick]  (d412.center) edge (d47.center);
\draw[very thick]  (d42.center) edge (d47.center);
\draw[very thick]  (d43.center) edge (d46.center);
\draw[very thick]  (d41.center) edge +(0,.4);
\draw[very thick]  (d41.center) edge +(-.4,0);
\draw[very thick]  (d44.center) edge +(0,.4);
\draw[very thick]  (d44.center) edge +(.4,0);
\draw[very thick]  (d45.center) edge +(0,-.4);
\draw[very thick]  (d45.center) edge +(.4,0);
\draw[very thick]  (d48.center) edge +(0,-.4);
\draw[very thick]  (d48.center) edge +(-.4,0);
\draw[very thick]  (d49.center) edge +(0,.4);
\draw[very thick]  (d49.center) edge +(-.4,0);
\draw[very thick]  (d410.center) edge +(0,.4);
\draw[very thick]  (d410.center) edge +(.4,0);
\draw[very thick]  (d411.center) edge +(0,-.4);
\draw[very thick]  (d411.center) edge +(.4,0);
\draw[very thick]  (d412.center) edge +(0,-.4);
\draw[very thick]  (d412.center) edge +(-.4,0);
\draw[very thick]  (d42.center) edge +(-.56,.5);
\draw[very thick]  (d42.center) edge +(-.46,.6);
\draw[very thick]  (d43.center) edge +(.56,.5);
\draw[very thick]  (d43.center) edge +(.46,.6);
\draw[very thick]  (d46.center) edge +(.56,-.5);
\draw[very thick]  (d46.center) edge +(.46,-.6);
\draw[very thick]  (d47.center) edge +(-.56,-.5);
\draw[very thick]  (d47.center) edge +(-.46,-.6);
\node[label=left:{$\color{BrickRed} \cdots$}] at (1.38,0) {};
\node[label=above:{$\color{BrickRed} \vdots$}] at (2.8,1.5) {};
\node[label=above:{$\color{BrickRed} \vdots$}] at (2.8,-2.4) {};
\node[label=right:{$\color{BrickRed} \cdots$}] at (4.3,0) {};
\node[label=above:{$\color{BrickRed} a_{i-1}$}] at (1.6,.31) {};
\node[label=above:{$\color{BrickRed} a_{i+1}$}] at (4.04,.31) {};
\node[label=below:{$\color{BrickRed} b_{i-1}$}] at (1.56,-.31) {};
\node[label=below:{$\color{BrickRed} b_{i+1}$}] at (4.04,-.31) {};
\node[label=above:{$\color{BrickRed} b_L^\prime$}] at (2.05,.86) {};
\node[label=above:{$\color{BrickRed} a_L^\prime$}] at (3.55,.86) {};
\node[label=below:{$\color{BrickRed} a_L^{\prime \prime}$}] at (2.05,-.86) {};
\node[label=below:{$\color{BrickRed} b_L^{\prime \prime}$}] at (3.55,-.86) {};
\node[label=left:{$\color{black!60} \ell_{i-1}$}] at (2.37,0) {};
\node[label=left:{$\color{black!60} \ell_i$}] at (3.2,0) {};
\node[label=left:{$\color{black!60} \ell_{i+1}$}] at (4.45,0) {};
\node[label=left:{$\color{black!60} \ell_L^\prime$}] at (3.24,1.05) {};
\node[label=left:{$\color{black!60} \ell_L^{\prime \prime}$}] at (3.24,-1.05) {};
\draw[rounded corners] (-4.7, -4.2) rectangle (-.3, 4.2) {};
\draw[rounded corners] (.5, -2.3) rectangle (5.1, 2.3) {};
\end{tikzpicture} 
\caption{Traintrack networks can be Feynman-parametrized iteratively, by splitting up each intersection into an `inner' traintrack and two `outer' traintracks. One first Feynman-parametrizes the outer traintracks in such a way that the resulting integrals only depend on ${\color{black!60}x_{\ell_i}}$ via the squared differences $({\color{black!60}\ell_i}, {\color{Cerulean} R_L^\prime})$ and $({\color{black!60}\ell_i}, {\color{Cerulean} R_L^{\prime \prime}})$. One can then Feynman-parametrize the inner traintrack by treating $({\color{Cerulean} R_L^\prime})$ and $({\color{Cerulean} R_L^{\prime \prime}})$ as normal external dual points.}
\label{fig:iterative_Feynman_paramatrization}
\end{figure}
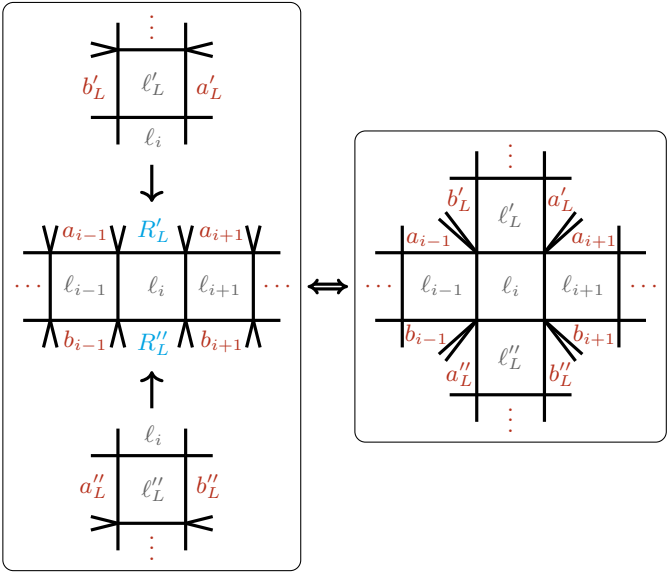

To derive an analogous $2L$-fold integral representation for any traintrack network, we strategically split the diagram into linear traintrack segments. In particular, wherever we encounter an intersecting pair of linear traintrack diagrams, we split the diagram into three parts---an `internal' traintrack comprised of all the loops along one axis (including the loop where the intersection occurs), and a pair of `external' traintracks formed by the loops to either side of this axis. This split is illustrated in Figure~\ref{fig:iterative_Feynman_paramatrization}. We then Feynman-parametrize the two outer traintrack diagrams separately, proceeding in a loop-by-loop fashion from the furthest-out loop towards the loop that borders the inner traintrack. In doing so, we treat ${\color{black!60}\ell_i}$ as an external dual point that is not being integrated over. This leaves us with an integral representation of each of these external traintracks in which the dual point ${\color{black!60}x_{\ell_i}}$ only appears in the last denominator factors---those corresponding to $({\color{Cerulean} R_L}, {\color{BrickRed} d})$ in equation~\eqref{eq:traintrack_parametric_form}, where $({\color{BrickRed} d})$ is now $({\color{black!60}\ell_i})$ and $({\color{Cerulean} R_L})$ is the last composite point that has been generated by the loop-by-loop integration procedure (which we have written as $({\color{Cerulean} R_L^\prime})$ or $({\color{Cerulean} R_L^{\prime \prime}})$ in Figure~\ref{fig:iterative_Feynman_paramatrization}). As a consequence, we can now Feynman-parametrize the inner traintrack diagram in the same manner by simply identifying $({\color{BrickRed} a_i}) = ({\color{Cerulean} R_L^\prime})$ and $({\color{BrickRed} b_i}) = ({\color{Cerulean} R_L^{\prime \prime}})$ in equation~\eqref{eq:traintrack_parametric_form}.

Carrying out this procedure iteratively, we can easily derive a $2L$-fold integral representation of any traintrack network, which will involve exactly $L{+}1$ denominator factors. The first $L$ factors take the form $\smash{( {\color{Cerulean} R_j^{(i)}}, {\color{Cerulean} R_j^{(i)}})}$, where $i$ indexes the linear traintrack segment each comes from, and $j$ runs over the number of loops $L^{(i)}$ in the $i^\text{th}$ traintrack segment. Note that our notation is such that $\smash{L= \sum_{i = 0}^T L^{(i)}}$, where we use $i=0$ to index the innermost traintrack and $T$ denotes the number of external traintrack segments (which will in general connect to each other in a nested fashion). We thus get an integral representation of the form
\begin{equation} \label{eq:feynman_parameter_form}
\int_0^\infty \frac{d^L {\color{ForestGreen} \alpha} \, d^L {\color{ForestGreen} \beta} \, d^L {\color{ForestGreen} \gamma}}{\text{GL}(1)^L}  \frac{\mathcal{N}}{\left[\prod^T_{i=0}\prod^{L^{(i)}}_{j=1}( {\color{Cerulean} R_j^{(i)}}, {\color{Cerulean} R_j^{(i)}}) \right] ( {\color{Cerulean} R^{(0)}_{L^{(0)}}}, {\color{BrickRed}  d^{(0)}})}\,,
\end{equation}
where $\smash{({\color{BrickRed}  d^{(0)}})}$ is the external dual point at the end of the innermost traintrack, and we have left the superscripts on the integration variables implicit. The dual points $\smash{( {\color{BrickRed} R_j^{(i)}} )} $ are defined recursively as before by 
\begin{equation}
 ( {\color{Cerulean} R_j^{(i)}} ) = {\color{ForestGreen} \gamma_j^{(i)}} ( {\color{Cerulean} R_{j-1}^{(i)}} ) + {\color{ForestGreen} \alpha_j^{(i)}} ({\color{BrickRed}  a_j^{(i)}}) + {\color{ForestGreen} \beta_j^{(i)}} ({\color{BrickRed} b_j^{(i)}}),
\end{equation}
where $\smash{({\color{Cerulean} R_0^{(i)}})=({\color{BrickRed} c^{(i)}})}$ is the external dual point at the beginning of each traintrack segement, and some of the points $\smash{({\color{BrickRed} a_j^{(i)}})}$ and $\smash{({\color{BrickRed} b_j^{(i)}})}$ should be replaced by the composite points $\smash{({\color{Cerulean} R_{L^{(k)}}^{(k)}})}$, as dictated by the topology of the diagram. Note that while this Feynman-parametrization procedure breaks the manifest dual conformal invariance of the original integral, this symmetry can be restored by rescaling the integration variables as done in~\cite{Bourjaily:2018ycu}; these integrals are therefore all functions of dual-conformal-invariant cross-ratios.

\vspace{.2cm}
\noindent {\bf Non-Polylogarithmicity}
\vspace{.1cm}

To diagnose the types of special functions traintrack networks give rise to, we sequentially compute residues of the integrand in~\eqref{eq:feynman_parameter_form}, thereby mimicking polylogarithmic integration, until an algebraic obstruction occurs. Following the method described in~\cite{Cao:2023tpx}, we do this by first solving for the vanishing locus of all the denominator factors $\smash{({\color{Cerulean} R_j^{(i)}},{\color{Cerulean} R_j^{(i)}})}$ in terms of the $\smash{{\color{ForestGreen} \gamma_j^{(i)}}}$ variables, using the fact that these factors are linear in $\smash{{\color{ForestGreen} \alpha_j^{(i)}}}$, $\smash{{\color{ForestGreen} \beta_j^{(i)}}}$, and $\smash{{\color{ForestGreen} \gamma_j^{(i)}}}$ when $\smash{({\color{Cerulean} R_{j-1}^{(i)}}, {\color{Cerulean} R_{j-1}^{(i)}})}$ also vanishes. Note that we can do this simultaneously for all $L$ factors of this type, regardless of whether they are associated with the innermost traintrack or one of the external ones. This allows us to sequentially compute $L$ residues with respect to the $\smash{{\color{ForestGreen} \gamma_j^{(i)}}}$ variables. 

Each of the $\smash{{\color{ForestGreen} \gamma_j^{(i)}}}$ residues gives rise to a Jacobian 
\begin{equation}
{\color{Violet} \mathcal{J}^{(i)}_j}={\color{ForestGreen} \alpha^{(i)}_j}({\color{BrickRed} a^{(i)}_j}, {\color{Cerulean} R^{(i)}_{j-1}})+{\color{ForestGreen} \beta^{(i)}_j}({\color{BrickRed} b^{(i)}_j}, {\color{Cerulean} R^{(i)}_{j-1}}) 
\end{equation}
in the denominator and entails making the replacements \\[-12pt]
\begin{equation}
(R^{(i)}_j,x)  \rightarrow \frac{{\color{ForestGreen} \vec{\alpha}^{(i)}_j} \cdot\mathbb{Q}^{(i)}_{j;x}\cdot {\color{ForestGreen} \vec{\alpha}^{(i)}_j}}{{\color{Violet}\mathcal{J}^{(i)}_j}}\,,
\label{eq:replacement}
\end{equation}
where ${\color{ForestGreen} \vec{\alpha}^{(i)}_j}=\{{\color{ForestGreen} \alpha^{(i)}_j},{\color{ForestGreen} \beta^{(i)}_j}\}$,
\begin{equation}
\mathbb{Q}^{(i)}_{j;x}=\begin{pmatrix}
(x,{\color{BrickRed} a^{(i)}_j})({\color{BrickRed}a^{(i)}_j}\!, {\color{Cerulean}R^{(i)}_{j-1}}) & \frac{1}{2}A({\color{BrickRed}a^{(i)}_j}\!,{\color{Cerulean}R^{(i)}_{j-1}},{\color{BrickRed}b^{(i)}_j}\!,x)\\[.2cm]
\frac{1}{2}A({\color{BrickRed}a^{(i)}_j}\!,{\color{Cerulean}R^{(i)}_{j-1}},{\color{BrickRed}b^{(i)}_j}\!,x) & (x,{\color{BrickRed}b^{(i)}_j})({\color{BrickRed}b^{(i)}_j}\!,{\color{Cerulean}R^{(i)}_{j-1}})
\end{pmatrix},
\label{eq:qmat}
\end{equation}
and $A(a,b,c,d)=(a,b)(c,d)+(a,d)(b,c)-(a,c)(b,d)$. After computing all $L$ residues, we are thus left with
\begin{equation} \label{eq:L_fold_residue}
\int_0^\infty \frac{d^L {\color{ForestGreen} \alpha} \, d^L {\color{ForestGreen} \beta} }{\text{GL}(1)^L}  \frac{\mathcal{N}}{\left[\prod^T_{i=0}\prod^{L_i}_{j=1}{\color{Violet}\mathcal{J}^{(i)}_j} \right] ({\color{Cerulean} R^{(0)}_{L^{(0)}}}, {\color{BrickRed} d^{(0)}})\big|_{\hookleftarrow}} \,,
\end{equation}
where $\hookleftarrow$ denotes the fact that we have iteratively made the replacements from equation~\eqref{eq:replacement} in $\smash{({\color{Cerulean} R^{(0)}_{L^{(0)}}}, {\color{BrickRed} d^{(0)}})}$.

In the case of the linear traintrack diagram, equation~\eqref{eq:L_fold_residue} can be further simplified. Namely---as shown in~\cite{Cao:2023tpx}---the Jacobian factors in the denominator will be canceled by factors that are generated when making the replacements described by equation~\eqref{eq:replacement}. The key fact is that each dual point $({\color{Cerulean} R_{i-1}})$ appears linearly in $({\color{Cerulean} R_L}, {\color{BrickRed} d})$, so precisely one inverse factor of each Jacobian is generated in the denominator. After all these Jacobian factors have cancelled out, there is just a single remaining denominator factor, which is quadratic in all extant Feynman parameters; computing a further residue in any of these variables thus results in an integral over a square root which is quartic in all Feynman parameters and which can be shown to describe a Calabi-Yau manifold.

In contrast, in full traintrack networks, whenever a new traintrack segment branches off, we end up replacing one of the external points $\smash{({\color{BrickRed} a_j^{(i)}})}$ or $\smash{({\color{BrickRed} b_j^{(i)}})}$ with a composite dual point $\smash{({\color{Cerulean} R_{L^{(k)}}^{(k)}}})$. These composite points then appear quadratically in the transformation~\eqref{eq:qmat} rather than linearly. As a result, additional factors of the Jacobian are generated in the numerator by these replacements. Moreover, as each of the factors $\smash{({\color{Cerulean} R_{j}^{(k)}},x)}$ will eventually be replaced by a polynomial that is quadratic in the $\smash{{\color{ForestGreen} \alpha_j^{(k)}}}$ and $\smash{{\color{ForestGreen} \beta_j^{(k)}}}$ parameters, the polynomial in the denominator ends up doubling in degree every time an external traintrack branching occurs. 

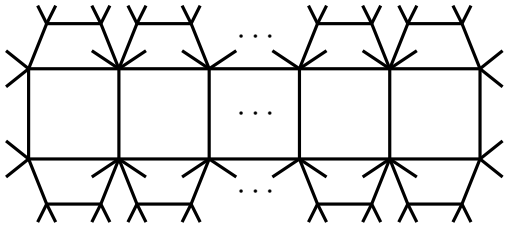
\begin{figure}
\centering
\begin{tikzpicture}[scale=1.2]
\node (d11) at (-4,.5) {};
\node (d12) at (-3,.5) {};
\node (d13) at (-2,.5) {};
\node (d14) at (-1,.5) {};
\node (d15) at (0,.5) {};
\node (d16) at (1,.5) {};
\node (d17) at (1,-.5) {};
\node (d18)at (0,-.5) {};
\node (d19) at (-1,-.5) {};
\node (d110) at (-2,-.5) {};
\node (d111) at (-3,-.5) {};
\node (d112) at (-4,-.5) {};
\node (dt1) at (-3.8,1) {};
\node (dt2) at (-3.2,1) {};
\node (dt3) at (-2.8,1) {};
\node (dt4) at (-2.2,1) {};
\node (dt5) at (-.8,1) {};
\node (dt6) at (-0.2,1) {};
\node (dt7) at (.2,1) {};
\node (dt8) at (.8,1) {};
\node (db1) at (-3.8,-1) {};
\node (db2) at (-3.2,-1) {};
\node (db3) at (-2.8,-1) {};
\node (db4) at (-2.2,-1) {};
\node (db5) at (-.8,-1) {};
\node (db6) at (-0.2,-1) {};
\node (db7) at (.2,-1) {};
\node (db8) at (.8,-1) {};
\draw[very thick]  (d11.center) edge (d12.center);
\draw[very thick]  (d12.center) edge (d13.center);
\draw[very thick]  (d13.center) edge (d14.center);
\draw[very thick]  (d14.center) edge (d15.center);
\draw[very thick]  (d15.center) edge (d16.center);
\draw[very thick]  (d16.center) edge (d17.center);
\draw[very thick]  (d17.center) edge (d18.center);
\draw[very thick]  (d18.center) edge (d19.center);
\draw[very thick]  (d19.center) edge (d110.center);
\draw[very thick]  (d110.center) edge (d111.center);
\draw[very thick]  (d111.center) edge (d112.center);
\draw[very thick]  (d112.center) edge (d11.center);
\draw[very thick]  (d11.center) edge (dt1.center);
\draw[very thick]  (dt1.center) edge (dt2.center);
\draw[very thick]  (dt2.center) edge (d12.center);
\draw[very thick]  (d12.center) edge (dt3.center);
\draw[very thick]  (dt3.center) edge (dt4.center);
\draw[very thick]  (dt4.center) edge (d13.center);
\draw[very thick]  (d14.center) edge (dt5.center);
\draw[very thick]  (dt5.center) edge (dt6.center);
\draw[very thick]  (dt6.center) edge (d15.center);
\draw[very thick]  (d15.center) edge (dt7.center);
\draw[very thick]  (dt7.center) edge (dt8.center);
\draw[very thick]  (dt8.center) edge (d16.center);
\draw[very thick]  (d112.center) edge (db1.center);
\draw[very thick]  (db1.center) edge (db2.center);
\draw[very thick]  (db2.center) edge (d111.center);
\draw[very thick]  (d111.center) edge (db3.center);
\draw[very thick]  (db3.center) edge (db4.center);
\draw[very thick]  (db4.center) edge (d110.center);
\draw[very thick]  (d19.center) edge (db5.center);
\draw[very thick]  (db5.center) edge (db6.center);
\draw[very thick]  (db6.center) edge (d18.center);
\draw[very thick]  (d18.center) edge (db7.center);
\draw[very thick]  (db7.center) edge (db8.center);
\draw[very thick]  (db8.center) edge (d17.center);
\draw[very thick]  (d12.center) edge (d111.center);
\draw[very thick]  (d13.center) edge (d110.center);
\draw[very thick]  (d14.center) edge (d19.center);
\draw[very thick]  (d15.center) edge (d18.center);
\draw[very thick]  (d11.center) edge +(-.25,.2);
\draw[very thick]  (d11.center) edge +(-.25,-.2);
\draw[very thick]  (d112.center) edge +(-.25,.2);
\draw[very thick]  (d112.center) edge +(-.25,-.2);
\draw[very thick]  (d16.center) edge +(.25,.2);
\draw[very thick]  (d16.center) edge +(.25,-.2);
\draw[very thick]  (d17.center) edge +(.25,.2);
\draw[very thick]  (d17.center) edge +(.25,-.2);
\draw[very thick]  (dt1.center) edge +(-.1,.2);
\draw[very thick]  (dt1.center) edge +(.1,.2);
\draw[very thick]  (dt2.center) edge +(-.1,.2);
\draw[very thick]  (dt2.center) edge +(.1,.2);
\draw[very thick]  (dt3.center) edge +(-.1,.2);
\draw[very thick]  (dt3.center) edge +(.1,.2);
\draw[very thick]  (dt4.center) edge +(-.1,.2);
\draw[very thick]  (dt4.center) edge +(.1,.2);
\draw[very thick]  (dt5.center) edge +(-.1,.2);
\draw[very thick]  (dt5.center) edge +(.1,.2);
\draw[very thick]  (dt6.center) edge +(-.1,.2);
\draw[very thick]  (dt6.center) edge +(.1,.2);
\draw[very thick]  (dt7.center) edge +(-.1,.2);
\draw[very thick]  (dt7.center) edge +(.1,.2);
\draw[very thick]  (dt8.center) edge +(-.1,.2);
\draw[very thick]  (dt8.center) edge +(.1,.2);
\draw[very thick]  (db1.center) edge +(-.1,-.2);
\draw[very thick]  (db1.center) edge +(.1,-.2);
\draw[very thick]  (db2.center) edge +(-.1,-.2);
\draw[very thick]  (db2.center) edge +(.1,-.2);
\draw[very thick]  (db3.center) edge +(-.1,-.2);
\draw[very thick]  (db3.center) edge +(.1,-.2);
\draw[very thick]  (db4.center) edge +(-.1,-.2);
\draw[very thick]  (db4.center) edge +(.1,-.2);
\draw[very thick]  (db5.center) edge +(-.1,-.2);
\draw[very thick]  (db5.center) edge +(.1,-.2);
\draw[very thick]  (db6.center) edge +(-.1,-.2);
\draw[very thick]  (db6.center) edge +(.1,-.2);
\draw[very thick]  (db7.center) edge +(-.1,-.2);
\draw[very thick]  (db7.center) edge +(.1,-.2);
\draw[very thick]  (db8.center) edge +(-.1,-.2);
\draw[very thick]  (db8.center) edge +(.1,-.2);
\draw[very thick]  (d12.center) edge +(-.3,.2);
\draw[very thick]  (d12.center) edge +(.3,.2);
\draw[very thick]  (d13.center) edge +(-.3,.2);
\draw[very thick]  (d13.center) edge +(.3,.2);
\draw[very thick]  (d14.center) edge +(-.3,.2);
\draw[very thick]  (d14.center) edge +(.3,.2);
\draw[very thick]  (d15.center) edge +(-.3,.2);
\draw[very thick]  (d15.center) edge +(.3,.2);
\draw[very thick]  (d18.center) edge +(-.3,-.2);
\draw[very thick]  (d18.center) edge +(.3,-.2);
\draw[very thick]  (d19.center) edge +(-.3,-.2);
\draw[very thick]  (d19.center) edge +(.3,-.2);
\draw[very thick]  (d110.center) edge +(-.3,-.2);
\draw[very thick]  (d110.center) edge +(.3,-.2);
\draw[very thick]  (d111.center) edge +(-.3,-.2);
\draw[very thick]  (d111.center) edge +(.3,-.2);
\node[label=center:{\large $\cdots$}] at (-1.46,0) {};
\node[label=center:{\large $\cdots$}] at (-1.46,.85) {};
\node[label=center:{\large $\cdots$}] at (-1.46,-.87) {};
\end{tikzpicture} 
\caption{The `caterpillar' diagrams, in which a single loop is attached the to the top and bottom of each internal traintrack loop.}
\label{fig:caterpillar_diagram}
\end{figure}

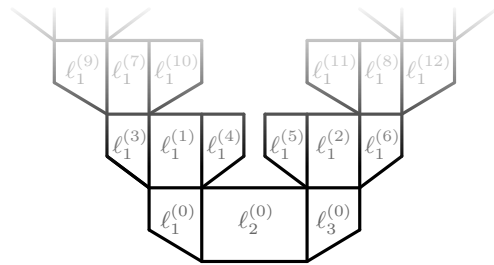
\begin{figure}[b]
\centering
\begin{tikzpicture}[scale=1.4]
\node[fill,circle,inner sep=.43] (a1) at (-.5,0.1) {};
\node[fill,circle,inner sep=.43] (a2) at (.5,0.1) {};
\node[fill,circle,inner sep=.43] (b1) at (-1,.4) {};
\node[fill,circle,inner sep=.43] (b2) at (1,.4) {};
\node[fill,circle,inner sep=.43] (c1) at (-1,.8) {};
\node[fill,circle,inner sep=.43] (c2) at (-.5,.8) {};
\node[fill,circle,inner sep=.43] (c3) at (.5,.8) {};
\node[fill,circle,inner sep=.43] (c4) at (1,.8) {};
\node[fill,circle,inner sep=.43] (d1) at (-1.4,1.1) {};
\node[fill,circle,inner sep=.43] (d2) at (-.1,1.1) {};
\node[fill,circle,inner sep=.43] (d3) at (.1,1.1) {};
\node[fill,circle,inner sep=.43] (d4) at (1.4,1.1) {};
\node[fill,circle,inner sep=.43] (e1) at (-1.4,1.5) {};
\node[fill,circle,inner sep=.43] (e2) at (-1,1.5) {};
\node[fill,circle,inner sep=.43] (e3) at (-.5,1.5) {};
\node[fill,circle,inner sep=.43] (e4) at (-.1,1.5) {};
\node[fill,circle,inner sep=.43] (e5) at (.1,1.5) {};
\node[fill,circle,inner sep=.43] (e6) at (.5,1.5) {};
\node[fill,circle,inner sep=.43] (e7) at (1,1.5) {};
\node[fill,circle,inner sep=.43] (e8) at (1.4,1.5) {};
\node[fill,circle,inner sep=.43] (f1) at (-1.9,1.8) {};
\node[fill,circle,inner sep=.43] (f2) at (-.5,1.8) {};
\node[fill,circle,inner sep=.43] (f3) at (.5,1.8) {};
\node[fill,circle,inner sep=.43] (f4) at (1.9,1.8) {};
\node[fill,circle,inner sep=.43] (g1) at (-1.9,2.2) {};
\node[fill,circle,inner sep=.43] (g2) at (-1.4,2.2) {};
\node[fill,circle,inner sep=.43] (g3) at (-1,2.2) {};
\node[fill,circle,inner sep=.43] (g4) at (-.5,2.2) {};
\node[fill,circle,inner sep=.43] (g5) at (.5,2.2) {};
\node[fill,circle,inner sep=.43] (g6) at (1,2.2) {};
\node[fill,circle,inner sep=.43] (g7) at (1.4,2.2) {};
\node[fill,circle,inner sep=.43] (g8) at (1.9,2.2) {};
\node[fill,circle,inner sep=.43] (h1) at (-2.3,2.5) {};
\node[fill,circle,inner sep=.43] (h2) at (-1,2.5) {};
\node[fill,circle,inner sep=.43] (h3) at (1,2.5) {};
\node[fill,circle,inner sep=.43] (h4) at (2.3,2.5) {};
\node[fill,circle,inner sep=.43] (i1) at (-1.9,2.5) {};
\node[fill,circle,inner sep=.43] (i2) at (-1.4,2.5) {};
\node[fill,circle,inner sep=.43] (i3) at (1.4,2.5) {};
\node[fill,circle,inner sep=.43] (i4) at (1.9,2.5) {};
\draw[very thick]  (a1.center) edge (a2.center);
\draw[very thick]  (b1.center) edge (a1.center);
\draw[very thick]  (b2.center) edge (a2.center);
\draw[very thick]  (c1.center) edge (b1.center);
\draw[very thick]  (c4.center) edge (b2.center);
\draw[very thick]  (c2.center) edge (a1.center);
\draw[very thick]  (c3.center) edge (a2.center);
\draw[very thick]  (c1.center) edge (c4.center);
\draw[very thick]  (d1.center) edge (c1.center);
\draw[very thick]  (d2.center) edge (c2.center);
\draw[very thick]  (d1.center) edge (c1.center);
\draw[very thick]  (d3.center) edge (c3.center);
\draw[very thick]  (d4.center) edge (c4.center);
\draw[very thick]  (e1.center) edge (d1.center);
\draw[very thick]  (e2.center) edge (c1.center);
\draw[very thick]  (e3.center) edge (c2.center);
\draw[very thick]  (e4.center) edge (d2.center);
\draw[very thick]  (e1.center) edge (e4.center);
\draw[very thick]  (e5.center) edge (d3.center);
\draw[very thick]  (e6.center) edge (c3.center);
\draw[very thick]  (e7.center) edge (c4.center);
\draw[very thick]  (e8.center) edge (d4.center);
\draw[very thick]  (e5.center) edge (e8.center);
\draw[very thick]  (e1.center) edge (f1.center);
\draw[very thick]  (e2.center) edge (f2.center);
\draw[very thick]  (e7.center) edge (f3.center);
\draw[very thick]  (e8.center) edge (f4.center);
\draw[very thick]  (g1.center) edge (f1.center);
\draw[very thick]  (g2.center) edge (e1.center);
\draw[very thick]  (g3.center) edge (e2.center);
\draw[very thick]  (g4.center) edge (f2.center);
\draw[very thick]  (g1.center) edge (g4.center);
\draw[very thick]  (g5.center) edge (f3.center);
\draw[very thick]  (g6.center) edge (e7.center);
\draw[very thick]  (g7.center) edge (e8.center);
\draw[very thick]  (g8.center) edge (f4.center);
\draw[very thick]  (g5.center) edge (g8.center);
\draw[very thick]  (g1.center) edge (h1.center);
\draw[very thick]  (g2.center) edge (h2.center);
\draw[very thick]  (g7.center) edge (h3.center);
\draw[very thick]  (g8.center) edge (h4.center);
\draw[very thick]  (g1.center) edge (i1.center);
\draw[very thick]  (g2.center) edge (i2.center);
\draw[very thick]  (g7.center) edge (i3.center);
\draw[very thick]  (g8.center) edge (i4.center);
\node[label=center:{\color{black!60} $\ell_{1}^{(0)}$}] at (-.72,0.5) {};
\node[label=center:{\color{black!60} $\ell_{2}^{(0)}$}] at (0.02,0.5) {};
\node[label=center:{\color{black!60} $\ell_{3}^{(0)}$}] at (0.76,0.5) {};
\node[label=center:{\color{black!60} $\ell_1^{(3)}$}] at (-1.176,1.22) {};
\node[label=center:{\color{black!60} $\ell_1^{(1)}$}] at (-.72,1.22) {};
\node[label=center:{\color{black!60} $\ell_1^{(4)}$}] at (-.28,1.22) {};
\node[label=center:{\color{black!60} $\ell_1^{(5)}$}] at (.32,1.22) {};
\node[label=center:{\color{black!60} $\ell_1^{(2)}$}] at (.76,1.22) {};
\node[label=center:{\color{black!60} $\ell_1^{(6)}$}] at (1.22,1.22) {};
\node[label=center:{\color{black!60} $\ell_1^{(9)}$}] at (-1.62,1.93) {};
\node[label=center:{\color{black!60} $\ell_1^{(7)}$}] at (-1.18,1.93) {};
\node[label=center:{\color{black!60} $\ell_1^{(10)}$}] at (-.73,1.93) {};
\node[label=center:{\color{black!60} $\ell_1^{(11)}$}] at (.77,1.93) {};
\node[label=center:{\color{black!60} $\ell_1^{(8)}$}] at (1.22,1.93) {};
\node[label=center:{\color{black!60} $\ell_1^{(12)}$}] at (1.66,1.93) {};
\fill[white,path fading=south] (-2.5,1) rectangle (2.5,2.55);
\end{tikzpicture} 
\caption{The `V formation' diagrams, in which traintracks are sequentially attached to the first available loop in the preceding external traintrack segment. We have suppressed all external momenta for graphical simplicity, but the momentum flowing into each vertex should be understood to be generic.}
\label{fig:V_formation_diagram}
\end{figure}

Take for example the `caterpillar' diagrams shown in Figure~\ref{fig:caterpillar_diagram}. After computing all the $\smash{{\color{ForestGreen} \gamma_j^{(i)}}}$ residues in one of these diagrams, one arrives at an integral of the form
\begin{equation} \label{eq:caterpillar_L_fold_residue}
\int_0^\infty \frac{d^L {\color{ForestGreen} \alpha} \, d^L {\color{ForestGreen} \beta} }{\text{GL}(1)^L}  \frac{\ \mathcal{N}\ \prod^{\frac23L}_{i=1} {\color{Violet}\mathcal{J}^{(i)}_1} }{P_c({\color{ForestGreen} \alpha},{\color{ForestGreen} \beta})} \,,
\end{equation}
where $P_c({\color{ForestGreen} \alpha},{\color{ForestGreen} \beta})$ is quadratic in the variables $\smash{{\color{ForestGreen} \alpha_j^{(0)}}}$ and $\smash{{\color{ForestGreen} \beta_j^{(0)}}}$, but quartic in all the $\smash{{\color{ForestGreen} \alpha_1^{(i)}}}$ and $\smash{{\color{ForestGreen} \beta_1^{(i)}}}$ variables. Or, as a second example, consider the `V formation' diagrams in Figure~\ref{fig:V_formation_diagram}. Since in this case each new traintrack is attached to the first available loop in the preceding traintrack segment, the corresponding Jacobian factors appear to increasingly high degree in the numerator. In particular, after computing all the $\smash{{\color{ForestGreen} \gamma_j^{(i)}}}$ residues one finds
\begin{equation} \label{eq:v_L_fold_residue}
\int_0^\infty \frac{d^L {\color{ForestGreen} \alpha} \, d^L {\color{ForestGreen} \beta} }{\text{GL}(1)^L}  \frac{\mathcal{N} \big( {\color{Violet}\mathcal{J}_1^\text{(1;2)}}\big) \big({\color{Violet} \mathcal{J}_1^\text{(3;6)}} \big)^2 \big( {\color{Violet}\mathcal{J}_1^\text{(7;8)}} \big)^3 \big({\color{Violet} \mathcal{J}_1^\text{(9;12)} }\big)^4 \cdots}{P_V({\color{ForestGreen} \alpha},{\color{ForestGreen} \beta})} \,,
\end{equation}
where we have adopted the abbreviated notation
\begin{equation}
{\color{Violet}\mathcal{J}_1^\text{(i;j)}} = {\color{Violet}\tilde{ \mathcal{J}}_1^\text{(i)} \tilde{\mathcal{J}}_1^\text{(i+1)} \cdots  \tilde{\mathcal{J}}_1^\text{(j-1)}  \tilde{\mathcal{J}}_1^\text{(j)}}\, ,
\end{equation}
and ${\color{Violet}\smash{\tilde{\mathcal{J}}^{(i)}_j}}$ represents the polynomial numerator left in ${\color{Violet}\smash{\mathcal{J}^{(i)}_j}}$ after the Jacobians associated with subsequent replacements have all been factored out into the denominator (see the appendix for further details on this notation). In addition, the order of $P_V({\color{ForestGreen} \alpha},{\color{ForestGreen} \beta})$ doubles every time a new traintrack segment is attached---so it is quadratic in the parameters associated with the innermost traintrack loop, quartic in the parameters associated with traintracks 1 and 2, octic in the parameters associated with traintracks 3, 4, 5, and 6, and so on.

The numerator factors that appear in these integrals pose a problem if we are interested in describing them with Calabi-Yau varieties. In order for a variety to be Calabi-Yau, there must exist a nowhere-vanishing holomorphic form~\cite{Hori:2003ic}. Any numerator factor ${\color{Violet}\smash{\mathcal{J}^{(i)}_j}}$ will vanish at some point in the integration domain, so the form we find for the leading singularity does not let us demonstrate this property.\footnote{Not coincidentally, these varieties also fail to satisfy the power-counting of a Calabi-Yau in the appropriate projective space, precisely because of the presence of the numerator.} While some of these numerator factors can drop in power when further residues are computed, many generically remain. As a result, we cannot using this approach associate the leading singularity of all traintrack networks with Calabi-Yau varieties.

However, there is a notable class of exceptions. The `cross' diagrams---namely, those traintrack networks that involve just a single traintrack intersection---only inherit numerator factors from the two external traintrack loops that border the inner traintrack. Fortuitously, if we compute a final residue with respect to one of the Feynman parameters that is associated with the loop where the traintracks intersect, these numerator factors cancel out and the degree of the remaining square root in the denominator is quartic in all remaining variables. As a result, the leading singularities of these diagrams can be cleanly associated with Calabi-Yau $(L{-}1)$-folds, as done for the linear traintracks. We provide further details of this calculation in an appendix.

\vspace{.2cm}
\noindent {\bf Differential Equations and Subtopologies}
\vspace{.1cm}

It has long been known that ladder-type Feynman diagrams such as the linear traintracks are related at adjacent loop orders by second-order differential operators~\cite{Drummond:2010cz}. In the simplest cases, these relations can be derived by acting with the Laplace operator on external dual points that appear in only a single propagator. More specifically, the Laplace operator $\square_{\color{BrickRed} a}$ that differentiates with respect to the dual coordinate ${\color{BrickRed} x^\mu_a}$ acts on this propagator as~\cite{Drummond:2006rz}
\begin{equation}
\square_{\color{BrickRed} a}  \frac{1}{({\color{BrickRed} a}, {\color{black!60}\ell})} = -4 i \pi^2 \delta^{(4)}({\color{BrickRed} x_a} - {\color{black!60}x_{\ell}}) \, ,
\end{equation}
which trivializes the integral over ${\color{black!60}x_\ell}$ by simply setting ${\color{black!60}x_{\ell}} = {\color{BrickRed} x_a}$. This operator thus supplies a relation between the original integral and one in which ${\color{black!60}x_{\ell}}$ has been traded for ${\color{BrickRed} x_a}$, effectively deleting the loop.

These types of inter-loop relations have previously been exploited to compute integrals that depend on the same set of external dual points at each loop order (see for instance~\cite{Drummond:2010cz,Drummond:2012bg,Caron-Huot:2018dsv,Henn:2018cdp,McLeod:2020dxg}), and the analogous relation that one gets by acting with $\square_{\color{BrickRed} d}$ on one end of the linear traintrack integral has also been leveraged at two loops~\cite{Kristensson:2021ani,Morales:2022csr}. However, we here underscore the fact that traintrack integrals satisfy a \emph{much larger} set of inter-loop relations. Namely, an independent relation of this type exists for \emph{every} external dual point that borders a traintrack network, as long as it appears in just a single propagator.

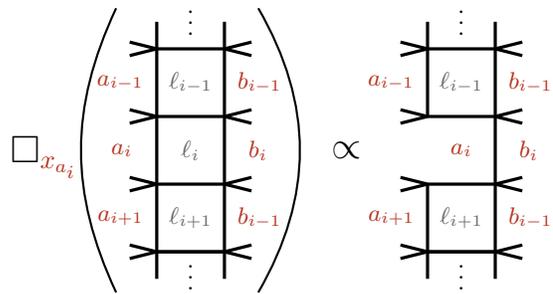
\begin{figure}
\centering
\begin{tikzpicture}[scale=0.9]
\node (d11) at (-2.5,1.5) {};
\node (d12) at (-2.5,.5) {};
\node (d13) at (-2.5,-.5) {};
\node (d14) at (-2.5,-1.5) {};
\node (d15) at (-1.5,-1.5) {};
\node (d16) at (-1.5,-.5) {};
\node (d17) at (-1.5,.5) {};
\node (d18) at (-1.5,1.5) {};
\draw[very thick]  (d11.center) edge (d12.center);
\draw[very thick]  (d12.center) edge (d13.center);
\draw[very thick]  (d13.center) edge (d14.center);
\draw[very thick]  (d14.center) edge (d15.center);
\draw[very thick]  (d15.center) edge (d16.center);
\draw[very thick]  (d16.center) edge (d17.center);
\draw[very thick]  (d17.center) edge (d18.center);
\draw[very thick]  (d18.center) edge (d11.center);
\draw[very thick]  (d12.center) edge (d17.center);
\draw[very thick]  (d13.center) edge (d16.center);
\draw[very thick]  (d14.center) edge (d15.center);
\draw[very thick]  (d11.center) edge +(0,.4);
\draw[very thick]  (d14.center) edge +(0,-.4);
\draw[very thick]  (d15.center) edge +(0,-.4);
\draw[very thick]  (d18.center) edge +(0,.4);
\draw[very thick]  (d11.center) edge +(-.4,.1);
\draw[very thick]  (d11.center) edge +(-.4,-.1);
\draw[very thick]  (d12.center) edge +(-.4,.1);
\draw[very thick]  (d12.center) edge +(-.4,-.1);
\draw[very thick]  (d13.center) edge +(-.4,.1);
\draw[very thick]  (d13.center) edge +(-.4,-.1);
\draw[very thick]  (d14.center) edge +(-.4,.1);
\draw[very thick]  (d14.center) edge +(-.4,-.1);
\draw[very thick]  (d15.center) edge +(.4,.1);
\draw[very thick]  (d15.center) edge +(.4,-.1);
\draw[very thick]  (d16.center) edge +(.4,.1);
\draw[very thick]  (d16.center) edge +(.4,-.1);
\draw[very thick]  (d17.center) edge +(.4,.1);
\draw[very thick]  (d17.center) edge +(.4,-.1);
\draw[very thick]  (d18.center) edge +(.4,.1);
\draw[very thick]  (d18.center) edge +(.4,-.1);
\node[label=center:{$\vdots$}] at (-2,2) {};
\node[label=center:{$\vdots$}] at (-2,-1.76) {};
\node[label=left:{$\color{BrickRed} a_{i-1}$}] at (-2.44,1) {};
\node[label=left:{$\color{BrickRed} a_i$}] at (-2.6,0) {};
\node[label=left:{$\color{BrickRed} a_{i+1}$}] at (-2.44,-1) {};
\node[label=center:{$\color{black!60} \ell_{i-1}$}] at (-2,1) {};
\node[label=center:{$\color{black!60} \ell_{i}$}] at (-2,0) {};
\node[label=center:{$\color{black!60} \ell_{i+1}$}] at (-2,-1) {};
\node[label=right:{$\color{BrickRed} b_{i-1}$}] at (-1.56,1) {};
\node[label=right:{$\color{BrickRed} b_i$}] at (-1.4,0) {};
\node[label=right:{$\color{BrickRed} b_{i-1}$}] at (-1.56,-1) {};
\draw [thick] (-3.1,-2.1) to [round left paren ] (-3.1,2.1);
\draw [thick] (-1,-2.1) to [round right paren] (-1,2.1);
\node[label=center:{\Large $\square_{\color{BrickRed} x_{a_i}}$}] at (-4.16,-0.1) {};    
\node[label=center:{\Large $\propto$}] at (.3,0) {};
\node (d21) at (1.5,1.5) {};
\node (d22) at (1.5,.5) {};
\node (d23) at (1.5,-.5) {};
\node (d24) at (1.5,-1.5) {};
\node (d25) at (2.5,-1.5) {};
\node (d26) at (2.5,-.5) {};
\node (d27) at (2.5,.5) {};
\node (d28) at (2.5,1.5) {};
\draw[very thick]  (d21.center) edge (d22.center);
\draw[very thick]  (d23.center) edge (d24.center);
\draw[very thick]  (d24.center) edge (d25.center);
\draw[very thick]  (d25.center) edge (d26.center);
\draw[very thick]  (d26.center) edge (d27.center);
\draw[very thick]  (d27.center) edge (d28.center);
\draw[very thick]  (d28.center) edge (d21.center);
\draw[very thick]  (d22.center) edge (d27.center);
\draw[very thick]  (d23.center) edge (d26.center);
\draw[very thick]  (d24.center) edge (d25.center);
\draw[very thick]  (d21.center) edge +(0,.4);
\draw[very thick]  (d24.center) edge +(0,-.4);
\draw[very thick]  (d25.center) edge +(0,-.4);
\draw[very thick]  (d28.center) edge +(0,.4);
\draw[very thick]  (d21.center) edge +(-.4,.1);
\draw[very thick]  (d21.center) edge +(-.4,-.1);
\draw[very thick]  (d22.center) edge +(-.4,.1);
\draw[very thick]  (d22.center) edge +(-.4,-.1);
\draw[very thick]  (d23.center) edge +(-.4,.1);
\draw[very thick]  (d23.center) edge +(-.4,-.1);
\draw[very thick]  (d24.center) edge +(-.4,.1);
\draw[very thick]  (d24.center) edge +(-.4,-.1);
\draw[very thick]  (d25.center) edge +(.4,.1);
\draw[very thick]  (d25.center) edge +(.4,-.1);
\draw[very thick]  (d26.center) edge +(.4,.1);
\draw[very thick]  (d26.center) edge +(.4,-.1);
\draw[very thick]  (d27.center) edge +(.4,.1);
\draw[very thick]  (d27.center) edge +(.4,-.1);
\draw[very thick]  (d28.center) edge +(.4,.1);
\draw[very thick]  (d28.center) edge +(.4,-.1);
\node[label=center:{$\vdots$}] at (2,2) {};
\node[label=center:{$\vdots$}] at (2,-1.76) {};
\node[label=left:{$\color{BrickRed} a_{i-1}$}] at (1.56,1) {};
\node[label=left:{$\color{BrickRed} a_{i+1}$}] at (1.56,-1) {};
\node[label=center:{$\color{black!60} \ell_{i-1}$}] at (2,1) {};
\node[label=center:{$\color{BrickRed} a_{i}$}] at (2,0) {};
\node[label=center:{$\color{black!60} \ell_{i+1}$}] at (2,-1) {};
\node[label=right:{$\color{BrickRed} b_{i-1}$}] at (2.44,1) {};
\node[label=right:{$\color{BrickRed} b_i$}] at (2.6,0) {};
\node[label=right:{$\color{BrickRed} b_{i-1}$}] at (2.44,-1) {};
\end{tikzpicture} 
\caption{One of the types of inter-loop relations that can be derived by acting with Laplacians on traintrack diagrams.}
\label{fig:differential_operator}
\end{figure}

Figure~\ref{fig:differential_operator} depicts one of the novel classes of relations that can be derived using this observation, when we differentiate with respect to a dual point that appear to the side of a linear traintrack segment. Interestingly, unlike previously-studied relations of this type, these differential operators relate the original integral to a factorized product of lower-loop traintrack integrals. Notably, these lower-loop integrals depend on complementary subsets of the external momenta, and can only give rise to special functions that are consistent with the corresponding lower-loop traintrack topologies. This tells us that these second-order differential operators can map the original traintrack integral to an object with much simpler analytic structure. 

Relations involving products of more than two lower-loop traintrack integrals can also be derived by acting on dual points that border loops where multiple traintracks come together---for instance, by acting on the dual point that borders the top of region $\smash{{\color{black!60}\ell_1^{(1)}}}$ in Figure~\ref{fig:V_formation_diagram}. In addition, we can use Lapacians such as $\square_{\color{BrickRed} a_L}$ or $\square_{\color{BrickRed} b_L}$ that act on the points near the end of linear traintrack segments to derive relations that resemble what one gets by acting with $\square_{\color{BrickRed} d}$, but in which the external momenta of the original diagram has been routed in a different way.

These differential relations concretely highlight the way in which traintrack networks involve integrals over the varieties associated with their subtopologies, which cannot be seen by studying just the leading singularities of these diagrams. In fact, already at two loops this nested structure may be discerned, insofar as the $\Delta_{2,2}$ coproduct of the two-loop elliptic double-box has been shown to take the form of a sum of over one-loop boxes with different choices of external legs, tensored with weight-two elliptic functions~\cite{Morales:2022csr}. Six of the terms in this sum correspond to the integrals one gets by applying Laplacian operators with respect to the six external dual points that appear at two loops. The fact that nine further boxes appear in this formula---which correspond to the groupings of external momenta one gets by contracting one edge in each loop---suggests that traintrack networks may even obey further classes of inter-loop differential equations, which cannot be derived using the Laplacian (as occurs in other examples~\cite{Drummond:2010cz}).

\vspace{.2cm}
\noindent {\bf Conclusions}
\vspace{.1cm}

In this letter, we have shown that the same iterative Feynman-parametrization procedure that was used to study the linear traintrack diagrams~\cite{Bourjaily:2018ycu} can be extended to the much larger class of traintrack networks. These integrals satisfy a vast web of inter-loop differential equations, and for generic kinematic configurations give rise to integrals over $(L{-}1)$-dimensional varieties. While it would also be interesting to understand the extent to which these varieties degenerate in massless and soft limits---such as limits in which all vertices are quadrivalent~\cite{Gurdogan:2015csr,Sieg:2016vap,Grabner:2017pgm}---we have left the study of these kinematic limits to future work.

It is not yet clear whether the varieties that appear in the leading singularities of traintrack networks are always Calabi-Yau. In order to better map out the types of special functions that can arise in the planar limit of massless quantum field theories, it would be highly desirable to show more conclusively when and if this property fails. If it does, it remains possible these integrals could be motivically Calabi-Yau in the sense of~\cite{Bonisch:2021yfw}. More generally, it would be interesting to connect the properties of these varieties directly to physical principles such as causality, locality, or unitarity as part of the recent resurgence of interest in Landau analysis~\cite{Dennen:2015bet,Dennen:2016mdk,Prlina:2017azl,Prlina:2017tvx,Prlina:2018ukf,Gurdogan:2020tip,Correia:2020xtr,Bourjaily:2020wvq,Collins:2020euz,Mizera:2021icv,Hannesdottir:2021kpd,Correia:2021etg,Yang:2022gko,Hannesdottir:2022bmo,Vergu:2020uur,Muhlbauer:2022ylo,He:2022tph,Bourjaily:2022vti,Mizera:2022dko,Lippstreu:2022bib,Hannesdottir:2022xki,Gardi:2022khw,Correia:2022dcu,Berghoff:2022mqu,Dlapa:2023cvx,Britto:2023rig,Lippstreu:2023oio}.

Finally, given the suggestive form taken by the $\Delta_{2,2}$ coproduct of the elliptic double box~\cite{Morales:2022csr}, it is natural to wonder whether traintrack networks satisfy inter-loop differential equations beyond those we have identified in this paper---and moreover, whether these differential equations might provide enough information to bootstrap higher-loop traintrack networks in the future (using methods analagous to those first developed for polylogarithmic integrals~\cite{Caron-Huot:2020bkp}). (We note in passing that the nested structure these differential operators expose bears some resemblance to the coaction principles observed elsewhere~\cite{Schlotterer:2012ny,Schnetz:2013hqa,Brown:2015fyf,Panzer:2016snt,Laporta:2017okg,Schnetz:2017bko,Caron-Huot:2019bsq,Gurdogan:2020ppd}.) Alternately, it might be possible to find a prescription for directly inverting these differential operators, as has been done for ladder integrals that depend on the same set of dual points at each loop order~\cite{Caron-Huot:2018dsv,Borinsky:2021gkd,Schnetz:2021ebf,Borinsky:2022lds}, or---even more ambitiously---to re-sum specific classes of traintrack networks in the coupling. In this way, by continuing to map out the landscape of perturbative quantum field theory, we hope to gain insight into the structure of perturbation theory at high orders, and eventually catch a glimpse of how---at least in conformal planar theories---it resums to give rise to nontrivial non-perturbative physics.

\acknowledgments

\vspace{.2cm}
\noindent {\bf Acknowledgments}
\vspace{.1cm}

We thank Claude Duhr, Cristian Vergu, and Chi Zhang for helpful discussions. The work of MvH was supported by the research grant 00025445 from Villum Fonden.

\vspace{.2cm}
\noindent {\bf Appendix: the Cross Diagrams}
\vspace{.1cm}

In this appendix we provide a more detailed analysis of the cross diagrams, which are constructed by intersecting only a single pair of linear traintrack diagrams. To show that the leading singularities of these diagrams encode Calabi-Yau manfiolds, we first consider the form they take after just $L^{(0)}$ residues have been computed with respect to the inner traintrack variables $\smash{\color{ForestGreen} \gamma_j^{(0)}}$. As with the linear traintrack diagrams, no numerator factors appear at this stage. However, the composite dual points $\smash{\color{Cerulean} (R^{(1)}_{L^{(1)}})}$ and $\smash{\color{Cerulean} (R^{(2)}_{L^{(2)}})}$ that arise in the innermost external traintack loops now appear in the denominator factor $\smash{({\color{Cerulean} R^{(0)}_{L^{(0)}}}, {\color{BrickRed} d^{(0)}})}$ due to the sequence of replacements described by~\eqref{eq:replacement}. In particular, if the external traintracks attach to loop $k$ of the inner traintrack, the replacement that must be carried out when computing the residue with respect to $\smash{\color{ForestGreen} \gamma_k^{(0)}}$ involves the matrix
\begin{widetext}
\begin{equation}
\mathbb{Q}^{(0)}_{k;x}=\begin{pmatrix}
\big(x,{\color{Cerulean} R^{(1)}_{L^{(1)}}}\big)\big({\color{Cerulean}R^{(1)}_{L^{(1)}}}, {\color{Cerulean}R^{(0)}_{k-1}}\big) & \frac{1}{2}A\big({\color{Cerulean}R^{(1)}_{L^{(1)}}},{\color{Cerulean}R^{(0)}_{k-1}},{\color{Cerulean}R^{(2)}_{L^{(2)}}},x\big)\\[.2cm]
\frac{1}{2}A\big({\color{Cerulean}R^{(1)}_{L^{(1)}}},{\color{Cerulean}R^{(0)}_{k-1}},{\color{Cerulean}R^{(2)}_{L^{(2)}}},x\big) & \big(x,{\color{Cerulean}R^{(2)}_{L^{(2)}}}\big)\big({\color{Cerulean}R^{(2)}_{L^{(2)}}},{\color{Cerulean}R^{(0)}_{k-1}}\big)
\end{pmatrix},
\label{eq:cross_central_replacement_mat}
\end{equation}
\end{widetext}
where we have simply made the identifications $\smash{({\color{BrickRed} a^{(0)}_j}) = ({\color{Cerulean} R^{(1)}_{L^{(1)}}})}$ and $\smash{({\color{BrickRed}b^{(0)}_j}) = ({\color{Cerulean}R^{(2)}_{L^{(2)}}})}$ in~\eqref{eq:qmat}. Importantly, this matrix is quadratic rather than linear with respect to the dual points $\smash{({\color{Cerulean} R^{(1)}_{L^{(1)}}})}$ and $\smash{({\color{Cerulean} R^{(2)}_{L^{(2)}}})}$. As a result, after the residues associated with the remaining ${\color{ForestGreen} \gamma_j^{(i)}}$ variables are computed, this matrix will end up taking the form
\begin{widetext}
\begin{equation} \label{eq:jacobian_denominators}
\mathbb{Q}^{(0)}_{j;x}=
\frac{1}{\big({\color{Violet} \mathcal{J}^{(1)}_{L^{(1)}}}\big)^2 \big({\color{Violet} \mathcal{J}^{(2)}_{L^{(2)}}}\big)^2}
\begin{pmatrix}
(\cdots) \big({\color{Violet} \mathcal{J}^{(2)}_{L^{(2)}}}\big)^2 & (\cdots) \big({\color{Violet} \mathcal{J}^{(1)}_{L^{(1)}}}\big) \big({\color{Violet} \mathcal{J}^{(2)}_{L^{(2)}}}\big) \\[.2cm]
(\cdots)\big({\color{Violet} \mathcal{J}^{(1)}_{L^{(1)}}}\big) \big({\color{Violet} \mathcal{J}^{(2)}_{L^{(2)}}}\big) &(\cdots)  \big({\color{Violet} \mathcal{J}^{(1)}_{L^{(1)}}}\big)^2
\end{pmatrix},
\end{equation}
\end{widetext}
where the suppressed expressions $(\cdots)$ are rational functions of the variables $\smash{\color{ForestGreen} \alpha_j^{(1)}}$, $\smash{\color{ForestGreen}\beta_j^{(1)}}$, $\smash{\color{ForestGreen}\alpha_j^{(2)}}$, and $\smash{\color{ForestGreen}\beta_j^{(2)}}$. As the residue operation with respect to $\smash{\color{ForestGreen} \gamma_k^{(0)}}$ itself only generates a single power of $\smash{\color{Violet} \mathcal{J}^{(1)}_{L^{(1)}} \mathcal{J}^{(2)}_{L^{(2)}}}$ in the denominator of the integrand, we end up with a net factor of $\smash{\color{Violet} \mathcal{J}^{(1)}_{L^{(1)}}\mathcal{J}^{(2)}_{L^{(2)}}}$ in the numerator. 

Although the Jacobians $\smash{\color{Violet} \mathcal{J}^{(1)}_{L^{(1)}}}$ and $\smash{\color{Violet}\mathcal{J}^{(2)}_{L^{(2)}}}$ both begin as polynomials, they end up becoming rational functions when further residue replacements are made. In particular, as $\smash{\color{Violet}\mathcal{J}^{(i)}_{L^{(i)}}}$ is linear with respect to $(\smash{\color{Cerulean} R^{(i)}_{L^{(i)}-1}})$, it will end up taking the form
\begin{equation} \label{eq:jacobian_replacements}
{\color{Violet} \mathcal{J}^{(i)}_{L^{(i)}}}= \frac{{\color{ForestGreen} \alpha^{(i)}_{L^{(i)}}}(\cdots )+{\color{ForestGreen} \beta^{(i)}_{L^{(i)}}}(\cdots) }{\color{Violet} \mathcal{J}_{L^{(i)}-1}^{(i)}\cdots \mathcal{J}_{0}^{(i)}}
\end{equation}
after all other $\smash{\color{ForestGreen} \gamma_{j}^{(i)}}$ residues are computed, where the suppressed expressions are $(\cdots)$ polynomials in the $\smash{\color{ForestGreen} \alpha_{j}^{(i)}}$ and $\smash{\color{ForestGreen} \beta_{j}^{(i)}}$ variables. The additional Jacobians that factor out into the denominator of $\smash{\color{Violet} \mathcal{J}^{(i)}_{L^{(i)}}}$ end up cancelling out entirely, to leave us with just the polynomial numerators that appear in~\eqref{eq:jacobian_replacements} in the numerator of the full integrand. In general, we denote the polynomial numerator that appears in $\smash{\color{Violet} \mathcal{J}^{(i)}_{j}}$ after all of these additional Jacobians have been factored out by $\smash{\color{Violet} \tilde{\mathcal{J}}^{(i)}_{j}}$. Thus, after the dust has settled, the integrals we are left with take the form 
\begin{equation}
\int_0^\infty \frac{d^L {\color{ForestGreen} \alpha} \, d^L {\color{ForestGreen} \beta }}{\text{GL}(1)^L}  \frac{\mathcal{N} \, {\color{Violet} \tilde{\mathcal{J}}^{(1)}_{L_1} \tilde{\mathcal{J}}^{(2)}_{L_2} }}{ P_{+}({\color{ForestGreen} \alpha}, {\color{ForestGreen} \beta}) }\,.
\end{equation}
where $P_+({\color{ForestGreen} \alpha}, {\color{ForestGreen} \beta})$ is quadratic in the variables $\smash{{\color{ForestGreen} \alpha_j^{(0)}}}$ and $\smash{{\color{ForestGreen} \beta_j^{(0)}}}$, but quartic in the $\smash{{\color{ForestGreen} \alpha_j^{(1)}}}$, $\smash{{\color{ForestGreen} \beta_j^{(1)}}}$, $\smash{{\color{ForestGreen} \alpha_j^{(2)}}}$, and $\smash{{\color{ForestGreen} \beta_j^{(2)}}}$ variables.

Finally, we can compute one more residue with respect to one of the $\smash{{\color{ForestGreen} \alpha_j^{(0)}}}$ or $\smash{{\color{ForestGreen} \beta_j^{(0)}}}$ variables. In order to make the Calabi-Yau property manifest, it turns out to be convenient to compute this last residue with respect to one of the Feynman parameters associated with the loop where the traintracks intersect, namely $\smash{\color{ForestGreen} \alpha^{(0)}_k}$ or $\smash{\color{ForestGreen}\beta^{(0)}_k}$, while deprojectivizing the other by setting it to 1. After doing this, we obtain a square root in the denominator, which is nothing more than the discriminant of the quartic polynomial $\smash{{\color{ForestGreen} \vec{\alpha}^{(0)}_k}\cdot \tilde{\mathbb{Q}}^{(0)}_{j;x}\cdot {\color{ForestGreen} \vec{\alpha}^{(0)}_k}}$, where $\smash{\tilde{\mathbb{Q}}^{(0)}_{j;x}}$ is just the polynomial numerator part of ${\mathbb{Q}}^{(0)}_{j;x}$ that remains after all Jacobians that can be factored out into the denominator have been. As can be seen from equation~\eqref{eq:jacobian_denominators} (after all Jacobian denominators have been pulled out), this discriminant will involve an overall factor of $\smash{\color{Violet} (\tilde{\mathcal{J}}^{(1)}_{L^{(1)}}\tilde{\mathcal{J}}^{(2)}_{L^{(2)}})^2}$, which can be pulled out of the square root to cancel the factor of $\smash{\color{Violet} \tilde{\mathcal{J}}^{(1)}_{L^{(1)}}\tilde{\mathcal{J}}^{(2)}_{L^{(2)}}}$ in the numerator. This finally lands us on a leading singularity of the form
\begin{equation}
\textbf{\textrm{LS}}_{\text{cross}}=\int_0^\infty \frac{d^{L-1} {\color{ForestGreen} \alpha} \, d^{L-1} {\color{ForestGreen} \beta} }{\text{GL}(1)^{L-1}}\frac{\mathcal{N}}{\sqrt{\mathcal{Q}_+({\color{ForestGreen} \alpha},{\color{ForestGreen} \beta})}}\,,
\end{equation}
where $\mathcal{Q}_+({\color{ForestGreen} \alpha},{\color{ForestGreen} \beta})$ is quartic in each of the remaining $L{-}1$ Feynman parameters, and has an overall degree of $4(L{-}1)$. Thus, this leading singularity can be cleanly associated with a Calabi-Yau $(L{-}1)$-fold, as done for the linear traintrack diagrams.

\bibliographystyle{utphys}
\bibliography{intersecting_traintracks}

\end{document}